\documentclass[conference]{IEEEtran}
\IEEEoverridecommandlockouts

\usepackage{cite}
\usepackage{amsmath,amssymb,amsfonts}
\usepackage{amsthm}
\usepackage{algorithm}
\usepackage{algpseudocode}
\usepackage{graphicx}
\usepackage{textcomp}
\usepackage{xcolor}
\usepackage{wrapfig}
\usepackage{hyperref}
\usepackage{booktabs}
\usepackage{array}
\usepackage[caption=false,font=footnotesize]{subfig}
\usepackage{bm} 
\def\BibTeX{{\rm B\kern-.05em{\sc i\kern-.025em b}\kern-.08em
    T\kern-.1667em\lower.7ex\hbox{E}\kern-.125emX}}

\theoremstyle{definition}

\begin{document}

\title{Stability and Performance of Online Feedback Optimization for Distribution Grid Flexibility}

\author{
    \IEEEauthorblockN{
        Florian Klein-Helmkamp\IEEEauthorrefmark{1},
        Tina Möllemann\IEEEauthorrefmark{1},
        Irina Zettl\IEEEauthorrefmark{1},
        and Andreas Ulbig\IEEEauthorrefmark{1}\IEEEauthorrefmark{2}\\}
    \IEEEauthorblockA{
        \IEEEauthorrefmark{1}IAEW at RWTH Aachen University, Aachen, Germany \\
        \IEEEauthorrefmark{2}Fraunhofer Center Digital Energy, Fraunhofer FIT, Aachen, Germany
        \\\{f.klein-helmkamp,i.zettl,a.ulbig\}@iaew.rwth-aachen.de
        \\ tina.moellemann@rwth-aachen.de
    }}

\maketitle

\begin{abstract}
The integration of distributed energy resources (DERs) into sub-transmission systems has enabled new opportunities for flexibility provision in ancillary services such as frequency and voltage support, as well as congestion management. This paper investigates the stability and performance of Online Feedback Optimization (OFO) controllers in ensuring reliable flexibility provision. A hierarchical control architecture is proposed, emphasizing safe transitions between system states within the Feasible Operating Region (FOR). We evaluate the controller's stability and performance through simulations of transitions to the vertices of the FOR, analyzing the impact of tuning parameters. The study demonstrates that controller stability is sensitive to parameter tuning, particularly gain and sensitivity approximations. Results demonstrate that improper tuning can lead to oscillatory or unstable behavior, highlighting the need for systematic parameter selection to ensure reliable operation across the full flexibility range.
\end{abstract}

\begin{IEEEkeywords}
Ancillary services, curative system operation, flexibility, online feedback optimization, system stability.
\end{IEEEkeywords}

\section{Introduction}
\begin{figure}[tb]
	\centering
	\includegraphics[width=\linewidth]{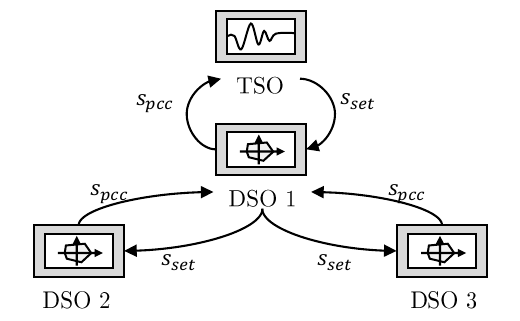}
	\caption{Cascaded interaction of system operators tracking set points for flexibility provision.}
	\label{fig:tso_dso}
\end{figure}

The increasing penetration of distributed energy resources (DER) at the sub-transmission level enables the provision of flexibility in power system operations. Potential use cases include ancillary services such as frequency and voltage support, congestion management, and portfolio optimization in response to market signals~\cite{Karagiannopoulos_2020, Stanojev_2021, Vagropoulos_2022}. The fulfillment of flexibility requests requires coordination of DER, with possible system architectures ranging from central coordination (e.g., at the control room level) to distributed approaches (e.g., at substation or plant levels). Flexibility can also extend across system boundaries, such as between transmission and distribution systems or neighboring grids (see \autoref{fig:tso_dso}), necessitating effective coordination and communication between system operators. This cross-boundary coordination introduces additional challenges related to interoperability, data sharing, and real-time communication protocols~\cite{Perez_2023}.

Regardless of the specific use case or coordination architecture, the provision of flexibility in online grid operation requires reliable and stable controllers. In recent years, there has been growing interest in online optimization for power system control~\cite{Häberle_2020}. These approaches rely on direct measurements of the system state as feedback, bypassing the need for explicit grid models. This reduces computational complexity while robustly steering the flexibility-providing system toward an optimal state. The characteristics of Online Feedback Optimization (OFO) make it a viable approach for addressing the flexibility dispatch problem, offering significant advantages over conventional methods such as feedforward optimization based on Optimal Power Flow (OPF).

\subsection{Related Work}
The dispatch problem of flexible actors in power systems, particularly in distribution systems, is commonly formulated as an Optimal Power Flow (OPF) problem. This formulation targets a requested operating point while satisfying grid constraints such as voltage limits and thermal asset loading~\cite{Früh_2022}. The mathematical representation of the OPF varies based on the specific use case and the technical characteristics of the flexibility-providing system~\cite{Givisiez_2020}. However, optimization-based approaches face significant challenges if applied to online system operation, particularly in distribution grids. First, these methods are computationally intensive due to the inclusion of nonlinear and nonconvex power flow equations as equality constraints. Second, they depend heavily on precise grid models and accurate estimates of system disturbances~\cite{Picallo_2022}. These challenges are particularly prohibitive for applications requiring rapid activation and provision of ancillary services within medium- to fast-time frames.

To address these limitations, online optimization approaches that operate directly on the physical grid have gained increased attention~\cite{Bernstein_2019}. Such methods employ optimization algorithms as closed-loop feedback controllers~\cite{Hauswirth_2016, Hauswirth_2017}. Relevant use cases in power systems include remedial actions at the transmission system level~\cite{Ortmann_2023} and voltage control at the distribution system level~\cite{Zhan_2024}. A notable example is tracking a load flow set point at the TSO-DSO interface~\cite{Klein-Helmkamp_2024} as part of the provision of distributed flexibility. However, the performance and robustness of such controllers are highly dependent on their specific tuning. Improper tuning can lead to instabilities or oscillatory behavior during flexibility provision~\cite{Ortmann_2024, Zettl_2024}.

Ensuring the stability of central controllers providing ancillary services is critical for maintaining the reliable and efficient online functionality of power systems. Instabilities can result in cascading failures, deviations from desired operating points, and violations of system constraints, ultimately compromising grid reliability. Therefore, the stability of feedback optimization-based controllers must be evaluated across the full operational flexibility range of the system. While theoretical stability guarantees exist~\cite{Colombino_2020, Bianchi_2024, Hauswirth_2024}, these results remain primarily theoretical. Importantly, tuning must ensure stability over the entire operational range to prevent instability or suboptimal performance. A comprehensive evaluation of controllers for online grid operation is essential to validate their practical reliability and robustness during flexibility provision.

\subsection{Main Contribution}
This paper proposes an Online Feedback Optimization (OFO)-based controller for flexibility provision in distribution grids and evaluates its stability across the feasible operating region $\mathcal{F}$. By simulating trajectory transitions between operating points at the interface between transmission and sub-transmission systems, we demonstrate that the controller's performance strongly depends on its parameter tuning. Specifically, we show that certain parameter sets can result in both stable and unstable trajectories within $\mathcal{F}$, emphasizing the critical need for systematic methods to ensure stable convergence for all theoretically feasible operating points under system constraints. These findings provide insights into the design and tuning of controllers for reliable flexibility provision in online power system operations.

\section{Control of Flexibility Provision}
In this section, we propose a controller for flexibility provision based on the Online Feedback Optimization (OFO) paradigm. Our approach focuses on ensuring feasible and stable transitions within the feasible operating region (FOR) of a power system while minimizing deviations from desired set points for complex load flow at the interface between TSO and DSO. We detail the formulation of the flexibility dispatch problem, the implementation of the OFO controller, and its stability analysis, highlighting key parameters such as controller gain and tuning matrices.

\subsection{Flexibility Dispatch}

\begin{figure}[tb]
	\centering
	\includegraphics[width=\linewidth]{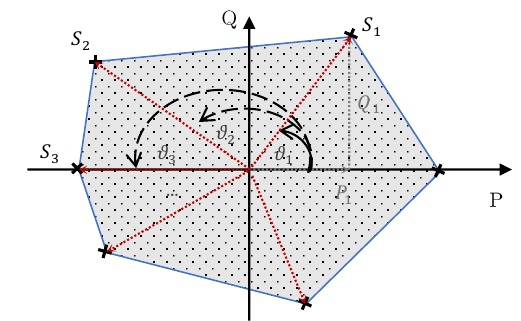}
	\caption{Feasible operating region $\mathcal{F}$ of a flexibility-providing system with vertices determined by a fixed angle $\vartheta$}
	\label{fig:for}
\end{figure}

In our approach to flexibility dispatch, we focus on the transition between two operating points within the feasible operating region (FOR) of a power system as shown in \autoref{fig:for}. The FOR is defined as the set of all load flows between two systems that satisfy both equality and inequality constraints of the flexibility-providing system, represented mathematically as:
\begin{equation}
\mathcal{F} = \{s_{pcc} \mid g(s_{pcc}, x) = 0, \ h(s_{pcc}, x) \leq 0\}
\end{equation}
where $ s_{pcc} $ denotes the complex power flow at the point of common coupling (PCC) between flexibility-providing and flexibility-requesting system, $ g $ represents equality constraints (e.g. the power flow equations), and $ h $ encompasses inequality constraints such as voltage bands, thermal limits of assets, and generator and load limits.
To facilitate flexibility dispatch, we formulate an optimization problem aimed at minimizing the deviation from a desired set point while adhering to operational constraints. This can be expressed as:
\begin{equation}
\label{eq:dispatch problem}
\begin{aligned}
    \min_{\mathbf{v}, \mathbf{s}, \mathbf{p}, \mathbf{q}} &\quad \Phi =  ||\mathbf{s}_{\text{set}} - \mathbf{s}_{\text{PCC}}||^{2}  \\
    \text{s.t.}
                &\quad \mathbf{v}_{\text{min}} \leq \mathbf{v} \leq \mathbf{v}_{\text{max}}, \\
                &\quad \mathbf{s}_{\text{min}} \leq \mathbf{s} \leq \mathbf{s}_{\text{max}}, \\
                &\quad \mathbf{p}_{\text{min}} \leq \mathbf{p} \leq \mathbf{p}_{\text{max}}, \\
                &\quad \mathbf{q}_{\text{min}} \leq \mathbf{q} \leq \mathbf{q}_{\text{max}}
\end{aligned}
\end{equation}
Here, $ \mathbf{s}_{\text{set}} = [p_{\text{set}}, q_{\text{set}}]^T $ is the target operating point that we aim to reach through flexibility dispatch. The optimization minimizes the Euclidean norm of the difference between measured apparent power flow $\mathbf{s}_{\text{pcc}} = [p_{\text{pcc}}, q_{\text{pcc}}]^T $ and this target set point. When transitioning between two operating points within the FOR, it is crucial to ensure that each intermediate and the final state remain feasible. Thus, at each iteration during our feedback optimization process, we evaluate whether the new state projections remain within the bounds defined by our FOR. We are thereby interpreting $\mathcal{F}$ as a two-dimensional projection of all system constraints onto the PQ-plane describing the load flow at the system boundaries (e.g. a coupling transformer). In this work we propose to solve the dispatch problem \eqref{eq:dispatch problem} in online operation on the physical system using the OFO paradigm. This allows us to neglect the non-convex and non-linear power flow equations in the optimization problem as the system state can be directly acquired by measurement, decreasing computational complexity while increasing robust against mismatch in model and physical plant. We detail this approach in the following subsection.

\subsection{Online Feedback Optimization}
\begin{figure}[tb]
    \centering
    \includegraphics[width=\linewidth]{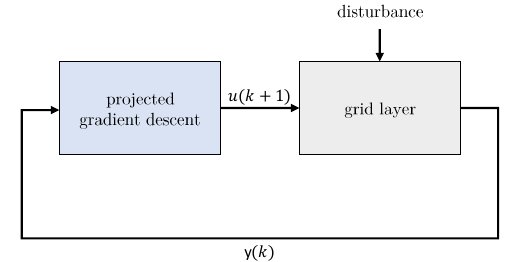}
    \caption{Projected gradient descent in closed loop with a physical grid layer subject to external disturbance}
    \label{fig:closed_loop}
\end{figure}

In this section we describe the proposed OFO controller in detail. Our approach is based on implementing projected gradient descent in closed-loop with the physical grid layer (see \autoref{fig:closed_loop}) described as a dynamic system with:
\begin{equation}
    y(k) = h(u) + d
\end{equation}
To this end we first approximate the system behavior by a function mapping control input to system output. This becomes possible by assuming time-scale separation between the actions of the OFO controller and all fast dynamics of the physical grid layer \cite{Hauswirth_2021}. Therefore, we can approximate the dynamic behavior of the system by its transition between steady state operating points $\nabla_u h(u, d)^T$. We express this input-output relationship in the optimization problem through the matrix \( H(u, d)^T \) defined as:
\begin{equation}
    H(u, d)^T = \begin{bmatrix} 
        I_n & \nabla_u h(u, d)^T 
    \end{bmatrix}.
\end{equation}

We define the vector of decision variables of the optimization problem describing the change in active and reactive power for the current iteration of OFO \( \mathbf{w} \in \mathbb{R}^p \) with:
\begin{equation}
    \mathbf{w} = \begin{bmatrix} 
        \Delta p \\ 
        \Delta q 
    \end{bmatrix}.
\end{equation}
The following quadratic programming (QP) problem describes the computation of the optimal control update \(\bm{\sigma}\) in the OFO framework. The objective minimizes the deviation from the desired gradient direction, weighted by the positive definite matrix \(G \succ 0\), which determines the relative importance of control adjustments. The matrices \(A \in \mathbb{R}^{m \times n}\) and \(C \in \mathbb{R}^{p \times n}\) define the input and output constraints, respectively, while \(\mathbf{b} \in \mathbb{R}^m\) and \(\mathbf{d} \in \mathbb{R}^p\) represent the corresponding constraint bounds. The control input \(\bm{u}(k)\) and system output \(\bm{y}(k)\) are updated iteratively, subject to these constraints and the sensitivity matrix \(\nabla h(\bm{u})\), ensuring feasible trajectories during the optimization process:
\begin{equation}
	\label{eq:qp}
    \begin{aligned}
        \bm{\sigma} = \arg \min_{\bm{w} \in \mathbb{R}^p} 
        & \|\bm{w} + G^{-1}H(\bm{u})^T \nabla \Phi(\bm{u}, \bm{y}) \|^2 \\
        \text{s.t.} \quad 
        & A (\bm{u}(k) + \alpha \bm{w}) \leq \bm{b}, \\
        & C (\bm{y}(k) + \alpha \nabla h(\bm{u}) \bm{w}) \leq \bm{d}.
    \end{aligned}
\end{equation}

For the control actions of iteration $k+1$ we scale the solution of the current step of gradient descent with a fixed parameter $\alpha$ and apply it to the previous control action with:
\begin{equation}
\label{eq:int_cont}
    \bm{u}(k+1) = \bm{u}(k) + \alpha \bm{\sigma}(u,y).
\end{equation}

A single iteration of the proposed OFO controller consists of the following steps driving the physical system to an optimal solution of the dispatch problem \eqref{eq:dispatch problem}:
\begin{enumerate}
	\item Measure the current system state \(\mathbf{y}(k)\), including power flows and bus voltages, to acquire real-time feedback from the grid.
	\item Compute the gradient of the cost function \(\nabla \Phi(\mathbf{u}, \mathbf{y})\) at the current step \(k\), where \(\Phi\) represents the deviation from the desired set point.
	\item Solve the quadratic programming problem in \eqref{eq:qp} to determine the updated control actions \(\mathbf{u}(k+1)\), ensuring feasibility with respect to the system constraints.
	\item Update the cost function \(\Phi(\mathbf{u}, \mathbf{y})\) based on the newly determined set point for load flow at the point of common coupling (PCC), \(\mathbf{s}_{\text{set}}\).
\end{enumerate}
The iterative adjustment of controllable flexibility within the respective grid layer drives the system to the optimal state $u^*$ in the sense of the dispatch problem \eqref{eq:dispatch problem}. To validate the controller behavior we analyze the resulting trajectory of system states.
\subsection{Tuning and Stability}

The stability and performance of the proposed OFO controller depend strongly on the tuning of key parameters, including the controller gain \(\alpha\), the sensitivity matrix \(\nabla h(\bm{u})\), and the weighting matrix \(G\). These parameters directly influence the controller's convergence behavior, robustness to system nonlinearities, and responsiveness during flexibility provision. In this subsection, we detail the role of these parameters and propose a set-based condition to evaluate the feasibility and stability of the controller. Furthermore we detail the simulative approach taken in this work. We begin by introducing a set-based condition to evaluate the feasibility of system states resulting from actuation by the OFO controller \cite{KH_2024}.

We define a trajectory set for an initial state $\mathbf{y}_{0}$ and a series of control inputs $\mathbf{u}(k) \in \mathcal{U}$ as the set of all possible resulting trajectories $\mathbf{y}(k)$ of a system with: 
\begin{equation}
\label{eq:def_oe}
\mathcal{E}(k) = \left\{ \mathbf{y}(k) \mid \mathbf{y}(k) = f(\mathbf{y}_0, \mathbf{u}(k), k), \ \mathbf{u}(k) \in \mathcal{U} \right\}
\end{equation}
Given a trajectory set for a series of control inputs $\mathcal{E}(k)$ we can evaluate it with respect to the FOR $\mathcal{F}$ of the system representing its physical and operational bounds. We assume a controller for a given tuning to be \emph{safe} if:
\begin{equation}
	\label{eq:safe_con}
    \mathcal{E} \subset \mathcal{F}
\end{equation}
holds. This allows us to simultaneously analyze the controller performance by evaluating the development of $\mathcal{E}(k)$ for an iteration $k$ and its stability and safety during convergence for its operational range.

In this paper we use the condition defined above to evaluate different parameter sets of the controller with respect to their performance and stability. We introduce the following relevant parameters with a brief description of their influence on system behavior during flexibility provision:

\subsubsection*{Controller Gain (\(\alpha\))}
The controller gain \(\alpha\) scales the step \(\bm{\sigma}(u,y)\) taken towards the solution of the optimization problem. The updated set point vector \(\bm{u}(k+1)\) is calculated iteratively according to \eqref{eq:int_cont} where \(\bm{\sigma}\) is the solution to the internal QP \eqref{eq:qp}. Careful tuning of \(\alpha\) is essential, as a larger \(\alpha\) increases the convergence speed, but may lead to oscillations if chosen excessively high.

\subsubsection*{Sensitivity Matrix (\(\nabla h(\bm{u})\))}
The sensitivity matrix \(\nabla h(\bm{u})\) captures the steady-state relationship between the input vector \(\bm{u}\) and the output vector \(\bm{y}\). It is defined as:
\begin{equation}
    \left[\nabla h(\bm{u})\right]_{i,j} = \frac{\partial h_i(\bm{u})}{\partial u_j}.
\end{equation}
The sensitivity matrix \(\nabla h(\bm{u})\) is derived by linearizing the dynamics of the system around a specific operating point \(\bm{u}_0\). This linearization introduces an error that grows with the deviation \(\|\bm{u} - \bm{u}_0\|\) due to the inherent non-linearities of the system. The controller gain \(\alpha\) scales the step size \(\alpha \bm{\sigma}\), and therefore there is a dependency between \(\alpha\) and the validity of the linearized model. Specifically, \(\alpha\) must be chosen small enough to bound the system output, despite the current error resulting from the linearization. Failure to satisfy this condition may lead to significant deviations in the output \(\bm{y}\) due to the inherit limitations of the linear approximation. As the stability of the system in presence of linearization errors in \(\nabla h(\bm{u})\) can be influenced by careful tuning of the controller gain, we are focusing in this work on the influence of the parameter $\alpha$ on controller performance.

\subsubsection*{Tuning Matrix (\(G\))}
The tuning matrix \(G\) weights the QP cost function, influencing the controller's convergence behavior:
\begin{equation}
    \begin{aligned}
        \|\bm{w} + G^{-1} H(\bm{u})^T \nabla \Phi(\bm{u}, \bm{y})\|^2_G
    \end{aligned}
\end{equation}
Lower values in \(G\) result in faster changes in the corresponding entries in \(\bm{w}\), allowing a faster operation in close proximity to the system constraints as available flexibility can be actuated faster. The interplay between \(\alpha\), \(\nabla h(\bm{u})\), and \(G\) in each iteration of the proposed controller determines the accuracy and speed of flexibility provision, as well as its stability during convergence to the targeted operating point. In order to guarantee a reliable, fast and stable adjustment to requested operating points the dynamic behavior during convergence of the proposed controller has to be evaluated for the full feasible operational range of the flexibility providing system. To achieve this, we step-wise target the vertices of the FOR of the system approximated by a polygon.

\section{Case Study}
This case study investigates the stability and performance of the proposed Online Feedback Optimization (OFO) controller during flexibility provision between an exemplary sub-transmission system and the superimposed transmission system. The analysis focuses on transitions to the vertices of the feasible operating region, highlighting the impact of different controller tunings on convergence behavior. By simulating the controller's operation under various parameter sets, we evaluate its stability, sensitivity to active constraints, and overall effectiveness in maintaining feasible states. This section also emphasizes the importance of tuning parameters such as gain and weighting matrix to ensure robust and reliable operation in dynamic grid scenarios.

\subsection{Scenario}
For this case study, we investigate the stability of flexibility provision from an exemplary sub-transmission system to the superimposed transmission system. We simulate and analyze the behavior of the proposed controller as it converges to the vertices of the feasible operating region (\(\mathcal{F}\)), as described in the previous section. \autoref{tab:scenario} summarizes the key parameters of the investigated scenario, including the total controllable flexibility present in the system~\cite{Meinecke_2020}. The system under study is a meshed high-voltage grid with a high penetration of wind power. 

To simplify the analysis, we assume that the proposed controller can directly actuate all distributed energy resources (DERs) and loads. Furthermore, we assume the system state is fully observable in each iteration of the controller, with no external disturbances present, to isolate the influence of internal tuning. The feasible operating region \(\mathcal{F}\) of the system is constrained by various operational limits. These include the upper voltage band of \(v_{\text{max}} = 1.1 \, \text{p.u.}\) for the first and second quadrants, as well as the capacity limits of the DERs in the second and third quadrants. The bounds in the third and fourth quadrants are predominantly determined by the limits of the coupling transformers, which result in the circular shape of the region shown in \autoref{fig:traj_008}. The FOR is approximated using 36 vertices corresponding to a resolution of \(10^\circ\). We therefore simulate 36 individual trajectories in each of the experiments conducted for this case study.

\begin{table}[tb]
	\centering
	\caption{Case Study: Grid Parameters \cite{Meinecke_2020}}
	\label{tab:scenario}
	\begin{tabular}{@{}llllll@{}}
		\toprule
		\textbf{$V_{\text{N}}$} & \textbf{Buses} & \textbf{Lines} & \textbf{Load} & \textbf{DER} & \textbf{Voltage Band} \\ \midrule
		110 kV & 119 & 151 & 535.3 MVA & 1432.25 MVA & $V_{\text{N}}\pm 10\%$  \\
		\bottomrule
	\end{tabular}
\end{table}

\subsection{Parameter Tuning}
We analyze the polygon approximation of the feasible operating region (\(\mathcal{F}\)) of the sub-transmission system, which is represented by 36 vertices corresponding to a resolution of \(10^\circ\). The edges of \(\mathcal{F}\) are defined by the active constraints that vary across different regions of the operating points \(\mathbf{s}_{\text{PCC}}\) on the PQ-plane. To evaluate the controller performance, we consider two parameter sets with different scalar gains \(\alpha\) and a fixed weighting matrix \(G\), as shown in \autoref{tab:ofo_tuning}.

\begin{table}[b]
	\centering
	\caption{Parameter tuning for OFO controller in case study.}
	\renewcommand{\arraystretch}{1.2}
	\setlength{\tabcolsep}{3pt} 
	\resizebox{\columnwidth}{!}{
		\begin{tabular}{@{}lccp{4.5cm}@{}}
			\toprule
			\textbf{Parameter Set} & \( \boldsymbol{\alpha} \) & \( \boldsymbol{G} \) & \textbf{Description} \\
			\midrule
			Set 1 & 0.008 & \( \mathbf{1} \) & Low \( \alpha \) for slower but smoother response. Stable convergence to vertices of $\mathcal{F}$. \\
			Set 2 & 0.02  & \( \mathbf{1} \) & Higher \( \alpha \) for larger steps toward $u^*$. Instable behavior observable. \\
			\bottomrule
		\end{tabular}%
	}
	\label{tab:ofo_tuning}
\end{table}

\begin{figure}[tb]
	\centering
	\includegraphics[width=\linewidth]{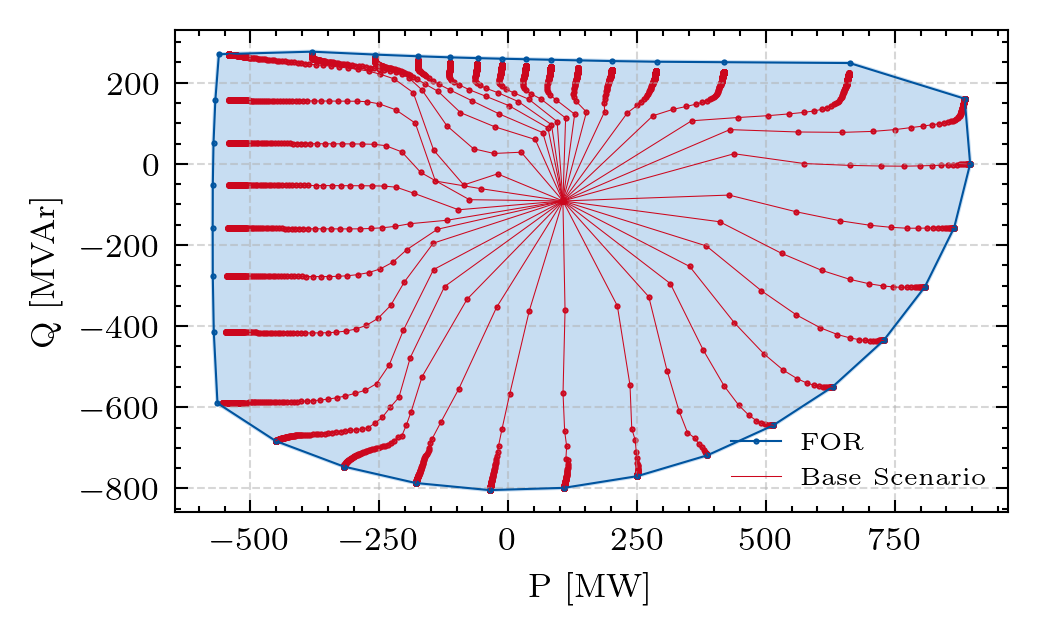}
	\caption{Stable trajectories of the proposed OFO controller for gain of \(\alpha~=~0.008\).}
	\label{fig:traj_008}
\end{figure}

We begin by analyzing the convergence behavior of the proposed controller for Parameter Set 1. With a conservatively small gain \(\alpha = 0.008\), the controller is expected to exhibit stable behavior, consistent with the theoretical guarantees of OF0. This is confirmed by the results shown in \autoref{fig:traj_008}, where stable convergence to all vertices of \(\mathcal{F}\) is observed. The performance of the controller near active constraints is influenced by the low gain, resulting in slower convergence rates. Additionally, differences in convergence speed are observed depending on the type of active constraint along the trajectories. For regions of $\mathcal{F}$ that are bounded by few individual constraints like the transformer limits at the PCC the convergence rate of the controller to $s_{\text{set}}$ is higher than for the upper region of $\mathcal{F}$ which is limited by the upper voltage band of the system. As numerous constraints become active the controller has to actuate more flexibility in order to ensure the feasibility of the resulting system states. This results in the slower convergence for the upper vertices of $\mathcal{F}$.

\begin{figure}[tb]
	\centering
	\includegraphics[width=\linewidth]{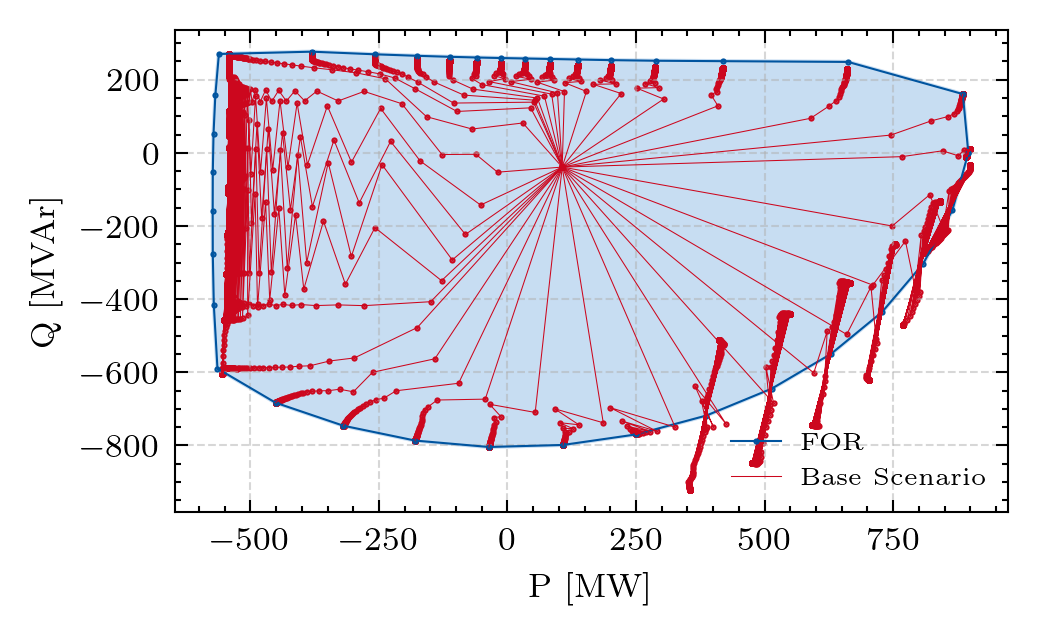}
	\caption{Partially stable convergence of trajectories for gain of \(\alpha = 0.02\).}
	\label{fig:traj_02}
\end{figure}

To contrast these results, we analyze the convergence for Parameter Set 2, characterized by a larger gain \(\alpha = 0.02\) (see \autoref{tab:ofo_tuning}). As shown in \autoref{fig:traj_02}, the controller takes larger steps toward the optimal solution of the dispatch problem (\(u^*\)). However, this results in intermediate system states that violate the constraints, as indicated by projections onto \(\mathcal{F}\) where \(y^*(k) \notin \mathcal{F}\). Moreover, oscillatory behavior is observed for many trajectories, particularly near regions of \(\mathcal{F}\) constrained by the coupling transformer limits. These oscillations can lead to critical system states, illustrating that parameter tuning significantly impacts stability and performance. The trajectories close to the left bound of the FOR are visibly not violating any system constraints, but nevertheless oscillate. This behavior is undesirable in grid operation because it leads to a continuous shift in operating points and impairs the provision of flexibility at the interface between systems. These findings underline the necessity of evaluating the full operational range of \(\mathcal{F}\) to identify parameter sets that ensure stable convergence for all use cases.

\begin{figure}[tb]
	\centering
	\includegraphics[width=\linewidth]{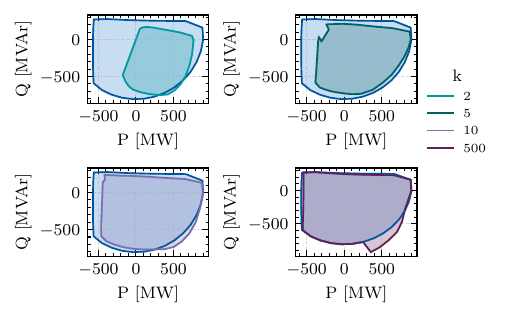}
	\caption{Trajectory sets \(\mathcal{E}(k)\) of the OFO controller with \(\alpha = 0.02\).}
	\label{fig:traj_sets}
\end{figure}

Finally, we perform a trajectory set evaluation to further analyze the controller's behavior. The increased gain in Parameter Set 2 improves the rate of convergence, as evidenced by the reachable regions within \(\mathcal{F}\) during the first two iterations of OFO. \autoref{fig:traj_sets} illustrates this, showing that around 52\% of \(\mathcal{F}\) is reachable within the first two iterations of the controller. For the exemplary iterations of $k \in \{2, 10, 500\}$ it is visible that $\mathcal{E} \not \subset \mathcal{F}$. Therefore, the controller tuning is unsafe in the sense of \eqref{eq:safe_con}. This highlights the trade-off between convergence speed and stability, emphasizing the need for systematic tuning approaches for the relevant parameters of the OFO controller.

\section{Conclusion}
In this paper, we analyzed the stability of a controller for flexibility provision in power systems based on a closed-loop implementation of projected gradient descent. Recognizing that appropriate controller performance is essential for reliable coordination of flexibility for ancillary services, we introduced the main tuning parameters influencing the behavior of the proposed Online Feedback Optimization (OFO) controller.

In our case study, we tested the OFO controller with two exemplary parameter sets, focusing on the gain $\alpha$. First, we demonstrated the converging behavior of the controller for a stable parameter set across the feasible operating region $\mathcal{F}$. Under the given scenario, the controller successfully reached a majority of $\mathcal{F}$ within several iterations. Second, we highlighted that for a different parameter set, the controller exhibited both stable and unstable behavior for various vertices of the FOR. This result emphasizes the importance of identifying parameter sets that ensure stable trajectories across the entire operational range of a flexibility-providing system, such as a sub-transmission system with controllable DER.

\section*{Acknowledgment}
\begin{wrapfigure}{r}{0.12\textwidth}
  \vspace{-\baselineskip}
  \vspace{-\baselineskip}
  \includegraphics[width=0.12\textwidth]{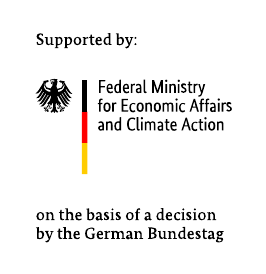}
  \vspace{-\baselineskip}
    \vspace{-\baselineskip}
        \vspace{-\baselineskip}
\end{wrapfigure}

This project received funding from the German Federal Ministry for Economic Affairs and Climate Action under the agreement no. 03EI4046E (PROGRESS).

\end{document}